# An $O(k^3 \log n)$-Approximation Algorithm for Vertex-Connectivity Survivable Network Design

Julia Chuzhoy[*]     Sanjeev Khanna[†]

October 16, 2018

## Abstract

In the Survivable Network Design problem (SNDP), we are given an undirected graph $G(V, E)$ with costs on edges, along with a connectivity requirement $r(u, v)$ for each pair $u, v$ of vertices. The goal is to find a minimum-cost subset $E^*$ of edges, that satisfies the given set of pairwise connectivity requirements. In the edge-connectivity version we need to ensure that there are $r(u, v)$ edge-disjoint paths for every pair $u, v$ of vertices, while in the vertex-connectivity version the paths are required to be vertex-disjoint. The edge-connectivity version of SNDP is known to have a 2-approximation. However, no non-trivial approximation algorithm has been known so far for the vertex version of SNDP, except for special cases of the problem. We present an extremely simple algorithm to achieve an $O(k^3 \log n)$-approximation for this problem, where $k$ denotes the maximum connectivity requirement, and $n$ denotes the number of vertices. We also give a simple proof of the recently discovered $O(k^2 \log n)$-approximation result for the single-source version of vertex-connectivity SNDP. We note that in both cases, our analysis in fact yields slightly better guarantees in that the $\log n$ term in the approximation guarantee can be replaced with a $\log \tau$ term where $\tau$ denotes the number of distinct vertices that participate in one or more pairs with a positive connectivity requirement.

---

[*]Toyota Technological Institute, Chicago, IL 60637. Email: cjulia@tti-c.org.

[†]Dept. of Computer & Information Science, University of Pennsylvania, Philadelphia, PA 19104. Email: sanjeev@cis.upenn.edu. Supported in part by a Guggenheim Fellowship, an IBM Faculty Award, and by NSF Award CCF-0635084.

# 1 Introduction

In the Survivable Network Design problem (SNDP), we are given an undirected graph $G(V, E)$ with costs on edges, and a connectivity requirement $r(u, v)$ for each pair $u, v$ of vertices. The goal is to find a minimum cost subset $E^*$ of edges, such that each pair $(u, v)$ of vertices is connected by $r(u, v)$ paths. In the edge-connectivity version (EC-SNDP), these paths are required to be edge-disjoint, while in the vertex-connectivity version (VC-SNDP), they need to be vertex-disjoint. It is not hard to show that EC-SNDP can be cast as a special case of VC-SNDP. We denote by $n$ the number of vertices in the graph and by $k$ the maximum pairwise connectivity requirement, that is, $\max_{u,v \in V} \{r(u, v)\}$. We also define a subset $T \subseteq V$ of vertices called *terminals*: a vertex $u \in T$ iff $r(u, v) > 0$ for some vertex $v \in V$.

The best current approximation algorithm for EC-SNDP is due to Jain [13], and it achieves a factor-2 approximation via the iterative rounding technique. At the same time no non-trivial approximation algorithms have been known for VC-SNDP, with the exception of several restricted special cases. Agrawal et. al. [1] showed a 2-approximation algorithm for the special case where maximum connectivity requirement $k = 1$. For $k = 2$, a 2-approximation algorithm was given by Fleischer [9]. The $k$-vertex connected spanning subgraph problem, a special case of VC-SNDP where for all $u, v \in V$ $r_{u,v} = k$, has been studied extensively. Cheriyan *et al.* [2, 3] gave an $O(\log k)$-approximation algorithm for this case when $k \leq \sqrt{n/6}$, and an $O(\sqrt{n/\epsilon})$-approximation algorithm for $k \leq (1 - \epsilon)n$. For large $k$, Kortsarz and Nutov [17] improved the preceding bound to an $O(\ln k \cdot \min\{\sqrt{k}, \frac{n}{n-k} \ln k\})$-approximation. Fakcharoenphol and Laekhanukit [8] improved it to an $O(\log n \log k)$-approximation, and further obtained an $O(\log^2 k)$-approximation for $k < n/2$. Very recently, Nutov [20] improved this to $O(\log k \cdot \log \frac{n}{n-k})$-approximation.

Kortsarz et. al. [15] showed that VC-SNDP is hard to approximate to within a factor of $2^{\log^{1-\epsilon} n}$ for any $\epsilon > 0$, when $k$ is polynomially large in $n$. This result was subsequently extended by Chakraborty et. al. [4] to a $k^\epsilon$-hardness for all $k > k_0$, where $k_0$ and $\epsilon$ are fixed positive constants. The existence of good approximation algorithms for small values of $k$ has remained an open problem, even for $k \geq 3$. In particular, when each connectivity requirement $r_{u,v} \in \{0, 3\}$, the best known approximation factor is polynomially large while only an APX-hardness is known on the hardness side.

A special case of VC-SNDP that has received much attention recently is the single-source version. In this problem there is a special vertex $s$ called the *source*, and all non-zero connectivity requirements involve $s$, that is, if $u \neq s$ and $v \neq s$, then $r(u, v) = 0$. Kortsarz et. al [15] showed that even this restricted special case of VC-SNDP is hard to approximate up to factor $\Omega(\log n)$, and recently Lando and Nutov [18] improved this to $(\log n)^{2-\epsilon}$-hardness of approximation for any constant $\epsilon > 0$. Both results only hold when $k$ is polynomially large in $n$. On the algorithmic side, Chakraborty et. al. [4] showed an $2^{O(k^2)} \log^4 n$-approximation for the problem. This result was later independently improved to $O(k^{O(k)} \log n)$-approximation by Chekuri and Korula [5], and to $O(k^2 \log n)$ by Chuzhoy and Khanna [7], and by Nutov [19]. Recently, Chekuri and Korula [6] simplified the analysis of the algorithm of [7]. We note that for the uniform case, where all non-zero connectivity requirements are $k$, Chuzhoy and Khanna [7] show a slightly better $O(k \log n)$-approximation algorithm, and the results of [6] extend to this special case.

A closely related problem to EC-SNDP and VC-SNDP is the element-connectivity SNDP. The input to the element-connectivity SNDP is the same as for EC-SNDP and VC-SNDP, and we also define the set $T \subseteq V$ of terminals as above. Given a problem instance, an *element* is any edge or any non-terminal vertex in the graph. We say that a pair $(s,t)$ of vertices is $k$-*element connected* iff for every subset $X$ of at most $(k-1)$ elements, $s$ and $t$ remain connected by a path when $X$ is removed from the graph. In other words, there are $k$ element-disjoint paths connecting $s$ to $t$; these paths are allowed to share terminals. Observe that if $(s,t)$ are $k$-vertex connected, then they are also $k$-element connected, and similarly, if $(s,t)$ are $k$-element connected, then they are also $k$-edge connected. The goal in the element-connectivity SNDP is to select a minimum-cost subset $E^*$ of edges, such that in the graph induced by $E^*$, each pair $(u,v)$ of vertices is $r(u,v)$-element connected. The element-connectivity SNDP was introduced in [14] as a problem of intermediate difficulty between edge-connectivity and vertex-connectivity, and the authors game a primal-dual $O(\log k)$-approximation for this problem. Subsequently, Fleischer *et al.* [10] gave a 2-approximation algorithm for element-connectivity SNDP via the iterative rounding technique, matching the 2-approximation guarantee of Jain [13] for EC-SNDP. We will use this result as a building block for our algorithm.

**Our results:** Our main result is as follows.

**Theorem 1** *There is a polynomial-time randomized $O(k^3 \log n)$-approximation algorithm for VC-SNDP, where $k$ denotes the largest pairwise connectivity requirement.*

In fact, our analysis gives a slightly better approximation guarantee of $O(k^3 \log |T|)$. The proof of this result is based on a randomized reduction that maps a given instance of VC-SNDP to a family of instances of element-connectivity SNDP. The reduction creates $O(k^3 \log n)$ instances, and has the property that any collection of edges that is feasible for *each one* of the element-connectivity SNDP instances generated above, is a feasible solution for the given VC-SNDP instance. We can thus use the known 2-approximation algorithm for element-connectivity SNDP to obtain the desired result.

We use these ideas to also give an alternative simple proof of the $O(k^2 \log n)$-approximation algorithm for the single-source VC-SNDP problem.

**Organization:** We present the proof of Theorem 1 in Section 2. Section 3 presents an alternative proof of the $O(k^2 \log n)$-approximation result for single-source VC-SNDP.

## 2 The Algorithm for VC-SNDP

Recall that in the VC-SNDP problem we are given an undirected graph $G(V, E)$ with costs on edges, and a connectivity requirement $r(u, v) \leq k$ for all $u, v \in V$. Additionally, we have a subset $T \subseteq V$ of terminals, and $r(u, v) > 0$ only if $u, v \in T$. Pairs of terminals with non-zero connectivity requirements are called *source-sink pairs*. We will use OPT to denote the cost of an optimal solution to the given VC-SNDP instance.

Our algorithm is as follows. We create $p$ copies of our original graph, say $G_1, G_2, \ldots, G_p$, where $p$ is a parameter to be determined later. For each copy $G_i$ we define a subset $T_i \subseteq T$ of terminals. We then view $G_i$ as an instance of element-connectivity SNDP, where the connectivity requirements are induced by the set $T_i$ of terminals as follows. For each $s, t \in T_i$ the new connectivity requirement is the same as the original one. For all other pairs the connectivity requirements are 0. Observe that for each $G_i$ the cost of an optimal solution for the induced element-connectivity SNDP instance is at most OPT. We then apply the 2-approximation algorithm of [10] to each one of the $p$ instances of $k$-element connectivity problem. Let $E_i$ denote the set of edges output by the 2-approximation algorithm on the instance defined on the $G_i$. Our final solution is $E^* = E_1 \cup E_2 \cup \ldots \cup E_p$. Clearly, the cost of the solution is at most $2p \cdot$ OPT. The main idea of our algorithm is that with the appropriate assignment of terminals to subsets $T_i$, the algorithm is guaranteed to produce a feasible solution.

**Definition 2.1** *Let $\mathcal{M}$ be the input collection of source-sink pairs and $T$ is the corresponding collection of terminals. We say that a family $\{T_1, \ldots, T_p\}$ of subsets of $T$ is* good *iff for each source-sink pair $(s, t) \in \mathcal{M}$, for each subset $X \subseteq T$ of size at most $(k-1)$, there is a subset $T_i$, $1 \leq i \leq p$, such that $s, t \in T_i$ and $X \cap T_i = \emptyset$.*

We show below that a good family of subsets exists for $p = O(k^3 \log n)$, and give a poly-time randomized algorithm to find such a family with high probability. We start by proving that such a family guarantees that the algorithm produces a feasible solution.

**Theorem 2** *Let $\{T_1, \ldots, T_p\}$ be a good family of subsets. Then the output $E^*$ of the above algorithm is a feasible solution to the VC-SNDP instance.*

*Proof.* Let $(s, t) \in \mathcal{M}$ be any source-sink pair, and let $X \subseteq V \setminus \{s, t\}$ be any collection of at most $(r(s, t) - 1) \leq (k - 1)$ vertices. It is enough to show that the removal of $X$ from the graph induced by $E^*$ does not separate $s$ from $t$. Let $X' = X \cap T$. Since $\{T_1, \ldots, T_p\}$ is a good family of subsets, there is some $T_i$ such that $s, t \in T_i$ while $T_i \cap X' = \emptyset$. Recall that set $E_i$ of edges defines a feasible solution to the element-connectivity SNDP instance corresponding to $T_i$. Then $X$ is a set of non-terminal vertices with respect to $T_i$. Since $s$ is $r(s, t)$-element connected to $t$ in the graph induced by $E_i$, the removal of $X$ from the graph does not disconnect $s$ from $t$. ∎

We now show how to find a good family of subsets $\{T_1, \ldots, T_p\}$. Let $p = 128k^3 \log n$, and set $q = p/(2k) = 64k^2 \log n$. Each terminal $t \in T$ selects uniformly at random $q$ indices from the set $\{1, 2, ..., p\}$ (repetitions are allowed). Let $\phi(t)$ denote the set of indices chosen by the terminal $t$. For each $1 \leq i \leq p$, we then define $T_i = \{t \mid i \in \phi(t)\}$.

**Theorem 3** *With high probability, the resulting family $\{T_1, \ldots, T_p\}$ of subsets is good.*

*Proof.* We extend the definition of $\phi()$ to an arbitrary subset $Z$ of vertices by defining $\phi(Z) = \bigcup_{t \in Z \cap T} \phi(t)$. Fix any source-sink pair $(s,t)$. Let $X$ be an arbitrary set of at most $(k-1)$ vertices that does not include $s, t$. Note that $|\phi(X)| \leq (k-1)q < p/2$. We say that the *bad event* $\mathcal{E}_1(s, t, X)$ *occurs* if $|\phi(s) \cap \phi(X)| \geq \frac{3q}{4}$. By Chernoff bounds,

$$\Pr[\mathcal{E}_1(s, t, X)] \leq e^{-q/32}.$$

We say that the *bad event* $\mathcal{E}_2(s, t, X)$ *occurs* if $\phi(s) \cap \phi(t) \subseteq \phi(X)$. We say that the set $X$ is a *bad set* for a pair $(s, t)$ if the event $\mathcal{E}_2(s, t, X)$ occurs. Note that if there is no bad set $X$ of size at most $(r(s,t) - 1)$ for every pair $(s,t) \in \mathcal{M}$, then $\{T_1, \ldots, T_p\}$ is a good family.

We observe that

$$\Pr[\mathcal{E}_2(s, t, X) \mid \overline{\mathcal{E}_1(s, t, X)}] \leq \left(1 - \frac{q/4}{p}\right)^q \leq e^{-q^2/4p} \leq e^{-q/8k}$$

Thus we can bound the probability of the event $\mathcal{E}_2(s, t, X)$ as follows:

$$\begin{aligned}
\Pr[\mathcal{E}_2(s, t, X)] &= \Pr[\mathcal{E}_2(s, t, X) \mid \mathcal{E}_1(s, t, X)]\Pr[\mathcal{E}_1(s, t, X)] + \Pr[\mathcal{E}_2(s, t, X) \mid \overline{\mathcal{E}_1(s, t, X)}]\Pr[\overline{\mathcal{E}_1(s, t, X)}] \\
&\leq \Pr[\mathcal{E}_1(s, t, X)] + \Pr[\mathcal{E}_2(s, t, X) \mid \overline{\mathcal{E}_1(s, t, X)}] \\
&\leq e^{-q/32} + e^{-q/8k} \\
&< n^{-4k}.
\end{aligned}$$

Hence, using the union bound, the probability that some bad set $X$ of size at most $(k-1)$ exists for any pair $(s, t)$ can be bounded by $n^{-2k}$. ∎

**Remark 1:** We note here that in the proof of Theorem 3, it suffices to ensure that the probability of the event $\mathcal{E}_2(s, t, X)$ is bounded by $|T|^{-4k}$ instead of $n^{-4k}$. To see this, observe that we need only to consider the sets $X$ that consist of terminal vertices. Moreover, the total number of source-sink pairs is bounded by $|T|^2$.

Combining Theorems 2 and 3 gives the following corollary:

**Corollary 1** *There is a randomized $O(k^3 \log n)$-approximation algorithm for VC-SNDP.*

**Remark 2:** We also note that this result implies that the integrality gap of the standard set-pair relaxation for VC-SNDP [12] has an integrality gap of $O(k^3 \log n)$. This follows from the fact that the 2-approximation result of [10] also establishes an upper bound of 2 on the integrality gap of the set-pair relaxation for element-connectivity. A lower bound of $\tilde{\Omega}(k^{1/3})$ is known on the integrality gap of the set-pair relaxation for VC-SNDP [4].

**Remark 3:** We notice that our algorithm carries over to the node-weighted version of VC-SNDP, and in particular an $\alpha$-approximation algorithm for the node-weighted element-connectivity SNDP would imply an $O(k^3 \alpha \log |T|)$-approximation for the node-weighted VC-SNDP.

## 3 The Algorithm for Single-Source VC-SNDP

In this section we show that an $O(k^2 \log n)$-approximation algorithm can be easily achieved using the above ideas for the single-source version of VC-SNDP. Several algorithms achieving similar approximation factors have been proposed recently [7, 6, 19]. While the algorithm and the analysis proposed here are elementary, we make use of the (relatively involved) 2-approximation algorithm of [10] as a black box. The algorithms of [7, 6] have the advantage that they are presented "from scratch", using only elementary tools, and when viewed as such they are rather simple.

The input to the single-source VC-SNDP is a graph $G = (V, E)$ with a special vertex $s$ called the source, and a subset $T$ of vertices called terminals. Additionally, for each $t \in T$ we are given a connectivity requirement $r(s, t) \leq k$. The goal is to select a minimum-cost subset $E' \subseteq E$ of edges, such that in the graph induced by $E'$ every terminal $t \in T$ is $r(s, t)$-vertex connected to $s$. This is clearly a special case of VC-SNDP, where the source-sink pairs are $\{(s, t)\}_{t \in T}$. As before, we create a family $\{T_1, \ldots, T_p\}$ of subsets of terminals, $T_i \subseteq T$ for all $1 \leq i \leq p$. We also create $p$ copies $G_1, \ldots, G_p$, and for each $G_i$ we solve the element-connectivity SNDP instance with connectivity requirements induced by terminals in $T_i$. Let $E_i$ be the 2-approximate solution to instance $G_i$. Our final solution is $E^* = \bigcup_{i=1}^{p} E_i$. Clearly, the cost of the solution is at most $2p(\text{OPT})$.

**Definition 3.1** *A family $\{T_1, \ldots, T_p\}$ of subsets of terminals is* good *iff for each terminal $t \in T$, for each subset $X \subseteq T$ of at most $(k-1)$ terminals, there is $T_i$ such that $t \in T_i$ and $T_i \cap X = \emptyset$.*

**Theorem 4** *If $\{T_1, \ldots, T_p\}$ is good family of subsets then the above algorithm produces a feasible solution.*

*Proof.* Let $t \in T$ and let $X \subseteq V \setminus \{s, t\}$ be any subset of at most $r(s, t) - 1 \leq (k - 1)$ vertices excluding $s$ and $t$. It is enough to prove that the removal of $X$ from the graph induced by $E^*$ does not disconnect $s$ from $t$. Let $X' = X \cap T$. Since $\{T_1, \ldots, T_p\}$ is a good family, there is some $T_i$ such that $t \in T_i$ and $T_i \cap X' = \emptyset$. Consider the solution $E_i$ to the corresponding $k$-element connectivity instance. Since vertices of $X$ are non-terminal vertices for the instance $G_i$, their removal from the graph induced by $E_i$ does not disconnect $s$ from $t$. ∎

Let $p = 4k^2 \log n$ and $q = p/(2k) = 2k \log n$. Each terminal $t \in T$ selects $q$ indices from the set $\{1, 2, ..., p\}$ uniformly at random with repetitions. Let $\phi(t)$ denote the set of indices chosen by the terminal $t$. For each $1 \leq i \leq p$, we then define $T_i = \{t \mid i \in \phi(t)\}$.

**Theorem 5** *With high probability, the resulting family of subsets $\{T_1, \ldots, T_p\}$ is good.*

*Proof.* Let $t \in T$ be any terminal and let $X$ be any subset of at most $r(s,t) - 1 \leq (k-1)$ terminals. As before, we extend the function $\phi$ to an arbitrary subset $Z$ of vertices by defining $\phi(Z) = \bigcup_{t \in Z \cap T} \phi(t)$. We say that *bad event $\mathcal{E}(t, X)$ occurs* iff $\phi(t) \subseteq \phi(X)$. The probability of $\mathcal{E}(t, X)$ is at most

$$\left(1 - \frac{kq}{p}\right)^q = \left(\frac{1}{2}\right)^q \leq n^{-2k}$$

Therefore, with high probability the event $\mathcal{E}(t, X)$ does not happen for any $t, X$ and then $\{T_1, \ldots, T_p\}$ is good. ∎

**Corollary 2** *There is a randomized $O(k^2 \log n)$-approximation algorithm for single-source VC-SNDP.*

## Acknowledgements

We thank Chandra Chekuri for his helpful comments on an earlier version of this paper.